# Atomic-Scale Strain Manipulation of a Charge Density Wave

Shang Gao[1], Felix Flicker[2,3], Raman Sankar[4], He Zhao[1], Zheng Ren[1], Bryan Rachmilowitz[1], Sidhika Balachandar[1], Fangcheng Chou[4], Kenneth Burch[1], Ziqiang Wang[1], Jasper Van Wezel[5] and Ilija Zeljkovic[1,¶]

[1]*Department of Physics, Boston College, Chestnut Hill, Massachusetts 02467, USA*  [2]*Department of Physics, UC Berkeley, Berkeley, California 94720*  [3] *Rudolf Peierls Centre for Theoretical Physics, University of Oxford, Oxford OX1 3NP*  [4]*Center for Condensed Matter Sciences, National Taiwan University, Taipei 10617, Taiwan*  [5]*Institute for Theoretical Physics, Institute of Physics, University of Amsterdam, 1090 GL Amsterdam, The Netherlands*
[¶] *Corresponding author* ilija.zeljkovic@bc.edu

## Abstract

A charge density wave (CDW) is one of the fundamental instabilities of the Fermi surface occurring in a wide range of quantum materials. In dimensions higher than one, where Fermi surface nesting can play only a limited role, the selection of the particular wave vector and geometry of an emerging CDW should in principle be susceptible to controllable manipulation. In this work, we implement a simple method for straining materials compatible with low-temperature scanning tunneling microscopy/spectroscopy (STM/S), and use it to strain-engineer new CDWs in *2H*-NbSe$_2$. Our STM/S measurements combined with theory reveal how small strain-induced changes in the electronic band structure and phonon dispersion lead to dramatic changes in the CDW ordering wave vector and geometry. Our work unveils the microscopic mechanism of a CDW formation in this system, and can serve as a general tool compatible with a range of spectroscopic techniques to engineer novel electronic states in any material where local strain or lattice symmetry breaking plays a role.

**Introduction**

Strain is one of few experimental handles available that can in principle be used to controllably and reversibly tune electronic and optical properties of materials, ranging from bulk [1–3] to reduced dimension materials [4–7]. However, achieving sufficient strain to generate novel behavior and simultaneously detecting the resulting emergent phenomena can be highly non-trivial. In thin films, strain has been successfully generated by utilizing the lattice mismatch between the film and the substrate, but the film growth on lattice mismatched substrates can often be challenging. In bulk single crystals, strain can be applied by attaching materials to piezoelectric substrates [1,2,8], but applicability to a wide range of characterization techniques has been limited by the necessity of independently controlling one or more piezoelectric stacks. Moreover, in real, imperfect materials, the strain may not transmit uniformly through the bulk to the top surface studied, so there is a pressing need for concomitant nanoscale structural and electronic characterization.

Transition metal dichalcogenides (TMDs) are an emerging family of extremely elastic quasi-2D materials able to withstand large amounts of in-plane strain (> 10%), thus providing the ideal playground for bandgap engineering, the design of new topological phases and the manipulation of many-body ground states [4,5]. A charge density wave (CDW) is one of the emergent states occurring in a range of TMDs [4], often accompanied by other, possibly competing, phases. A prototypical example is *2H*-NbSe$_2$, which exhibits both superconductivity ($T_c$ ~7.2 K) and a triangular (3Q) CDW phase ($T_{CDW}$ ~33 K) [9] that has intrigued the community for decades [10–22]. CDW formation can in principle arise from Fermi surface nesting, electron-electron interactions, or electron-phonon interactions [23]. Inspection of the Fermi surface of NbSe$_2$ shows little propensity to nesting [12], and alternative mechanisms have been sought since the earliest studies [24,25]. Although there is a growing consensus that electron-phonon coupling might play a role [15,19,20,26], a fundamental question remains as to what drives the choice of a particular CDW wavevector and geometry in this and other quasi-2D TMDs, and how these phases could be manipulated.

Here we implement a simple method that can achieve strain at the surface of a bulk material, while simultaneously allowing the measurement of electronic properties with atomic-scale precision. Our strain method exploits the mismatch in the thermal expansion coefficient (TEC) of materials to generate strain (Fig. 1a, Methods). Specifically, we glue a material of interest to a substrate with a vastly different TEC and cool it down from room temperature to ~4 K to induce strain. The striking simplicity of this method makes it suitable for rigid spatial constraints of spectroscopic imaging scanning tunneling microscopy (SI-STM) employed here, and it can also be easily extended to other low-temperature techniques. Although STM experiments have occasionally observed induced strain upon cooling down the sample [21,27], we note that our work is the first STM experiment to utilize the sample-substrate TEC mismatch for intentional strain application. Applying this method to *2H*-NbSe$_2$, we discover a remarkable emergence of two unexpected charge ordered phases, which we study to unveil the distinct roles of phonons and electrons in determining the ordering wavevector and geometry of a CDW.

## Results

Scanning tunneling microscopy (STM) topographs of the surface of unstrained NbSe$_2$ reveal a hexagonal lattice of Se atoms with a characteristic triangular (3Q) CDW ordering of ~$3a_0$ period (CDW-3a$_0$) below 33 K [10,21,28]. In our strained samples of *2H*-NbSe$_2$, in addition to detecting the well-known CDW-3a$_0$ in small patches (Fig. 1b), we reveal two additional types of charge ordering in other large regions of the sample – unidirectional "stripe" (1Q) ordering with 4a$_0$ period (CDW-4a$_0$) and a triangular (3Q) ordering with a 2a$_0$ period (CDW-2a$_0$) (Figs. 2(c,d)). The wavevectors of all observed CDWs are found to be oriented along the Γ-M directions, based on the Fourier transforms of STM topographs where each CDW peak lies exactly along the atomic Bragg wavevector $\mathbf{Q}_{Bragg}$ (Figs. 1(e-g)). We have observed the same CDW wavevectors on multiple NbSe$_2$ single crystals attached to substrates with mismatched TECs (see Methods). Interestingly, all the CDW wavevectors measured are commensurate with the lattice, in contrast to the recently observed incommensurate 1Q CDW phase with a ~3.5a$_0$ period, which was found in accidentally formed nanometer-scale "ribbons", and which could possibly be attributed to strain [21,29]. The magnitudes of the wavevectors identified in our experiments also does not change as a function of energy (SI Appendix, Section I), which eliminates a dispersive quasiparticle interference (QPI) signal [10] as the cause of our observations.

The presence of multiple distinct CDWs in different regions of the same strained single crystal suggests that these phases may be associated with strain of locally varying magnitude and/or direction. Although in an ideal homogeneous sample attached to a substrate under elastic deformation the strain is expected to remain laterally uniform as it is transmitted to the surface, this is unlikely to be the case in real materials that are inevitably inhomogeneous. In our NbSe$_2$ sample glued to a silica substrate by epoxy, inhomogeneous transmission of strain could arise due to the weak Van der Waals interlayer bonding that makes the material prone to warping [4] or inhomogeneous glue distribution at the interface. To shed light on what type of strain, if any, might play a role in the formation of each observed CDW, it is necessary to quantify strain at the atomic length scales. We start with an STM topograph $T(\mathbf{r})$ to which we apply the transformation $\mathbf{r} \rightarrow \mathbf{r} - \mathbf{u}(\mathbf{r})$ (where $\mathbf{u}(\mathbf{r})$ is the total displacement field obtained from the Lawler-Fujita algorithm [30]), such that the resulting topograph $T'(\mathbf{r} - \mathbf{u}(\mathbf{r}))$ contains a perfect hexagonal lattice. We disentangle the experimental artifacts (piezo and thermal drift) from structural strain in $\mathbf{u}(\mathbf{r})$ by fitting and subtracting a polynomial background to create the strain field $\mathbf{s}(\mathbf{r})$. The directional derivatives of $\mathbf{s}(\mathbf{r})$ form a strain tensor $s_{ij}(\mathbf{r}) \equiv \partial s_i(\mathbf{r})/\partial r_j$ (where $i, j = x, y$), and their linear combinations provide information on the strain type and magnitude [31–33] (SI Appendix, Section II). For example, we can extract biaxial (isotropic) strain as $(\mathbf{s}_{xx} + \mathbf{s}_{yy})/2$ (Fig. 2(c,d)). Although this algorithm cannot provide us with the absolute value of the applied strain, it can extract the relative local strain variations between different regions within a single STM topograph. Applying this procedure to the occasionally encountered boundaries between the CDW-3$a_0$, and the newly observed CDW-2$a_0$ and CDW-4$a_0$ phases (Figs. 2(a,b)), we find that regions hosting CDW-2$a_0$ and CDW-4$a_0$ are both under biaxial tensile strain (Figs. 2(c,d)) with a prominent uniaxial strain component relative to the CDW-3$a_0$ phase (SI Appendix, Section II). This is the first direct proof that in-plane tensile strain plays an important role in driving the new types of charge ordering.

To gain insight into the effects of strain on local electronic band structure in each region of the sample, we use QPI imaging, a method that applies two-dimensional Fourier transforms (FTs) to the STM $dI/dV$ maps to extract the electronic band dispersion. First, we focus on a large region of the sample hosting exclusively CDW-4$a_0$, in which the FTs of the $dI/dV$ maps show a circular QPI morphology (Fig. 3a-c) with the strongest intensity along the Γ-M direction. Higher momentum-space resolution of our data compared to previous experiments on NbSe$_2$ hosting a CDW-3$a_0$ [10] allows us to disentangle for the first time two distinct QPI peaks $\mathbf{Q}_1$ and $\mathbf{Q}_2$ (Fig. 3b), which arise from backscattering within the two Fermi surface pockets concentric around Γ (inset in Fig. 3b, SI Appendix, Section III). By measuring the positions of these peaks as a function of energy, we can map the two bands crossing the Fermi level along the Γ-M direction (Fig. 3d). Interestingly, the electronic band structure is only slightly different compared to that of the well-characterized unstrained material [10] (SI Appendix, Section IV), despite the dramatic changes in both the observed CDW wavelength and its geometry.

In the CDW-$2a_0$ region, we observe only the $Q_1$ vector, while $Q_2$ is notably absent in our measurable momentum range, in contrast to the CDW-$4a_0$ area (Fig. 3e-g). This suggests a more prominent change in the band structure. Our strain measurements in Fig. 2 reveal that this region of the sample is under tensile strain, which would lead to a larger momentum-space separation of the pockets around Γ (inset in Fig. 3f), owing to the concomitant increase in the interlayer tunneling (as the inter-layer orbital overlaps increase). Our QPI measurements however have been unable to detect any scattering vectors larger than $|Q_{Bragg}|/2$ in either CDW-$2a_0$ or CDW-$4a_0$ regions at any energy (SI Appendix, Section V), and we therefore cannot directly observe the shift of $Q_2$ to higher momenta. A possible explanation for the lack of signal at higher momenta may be canting of the orbital texture towards more in-plane orientations [34], making them less likely to be detected by the STM tip. Nevertheless, our measurements reveal that a larger distortion to the Fermi surface accompanies the formation of a CDW-$2a_0$.

**Discussion**

Having quantified the changes in the structural and electronic properties of regions hosting new CDWs, we turn to the fundamental question of what drives and stabilizes a particular CDW wavevector and geometry in this quasi-2D system. Taking into account the exactly commensurate nature of all observed CDWs, Fermi surface nesting is even more unlikely to play a role for the newly observed CDW phases. To provide further insight, we construct a simple model that captures the strain effects on both the electronic structure and phonon dispersion. We start with a tight-binding fit to the angle-resolved photoemission spectroscopy (ARPES) data [26,35], include the in-plane strain by modifying the hopping integrals and employ the Random Phase Approximation to calculate the resulting full electronic susceptibility $D_2(q)$ (Methods and SI Appendix, Section VI). We separately introduce the effect of the uniaxial strain on the phonons by shifting their bare energies differently in lattice-equivalent directions [29]. Within this model's description, the CDW ordering vector can be identified as the first wavevector for which the calculated susceptibility $D_2(q)$ exceeds the bare phonon energy $\Omega(q)$ identified in resonant inelastic X-ray scattering experiments [20,36].

In our model, we consider the effects of both uniaxial and biaxial in-plane strain, each modeled by a relative change in the nearest neighbor overlap integrals: σ associated with the uniaxial strain and $σ_i$ associated with the biaxial strain (for more details, see Methods and SI Appendix, Section VI). For simplicity, we explore the effects of the two types of strain separately. We find that biaxial strain by itself has very little effect on the shape of $D_2(\mathbf{q})$, while the uniaxial strain can lead to a significant change in $D_2(\mathbf{q})$ and induce different types of CDW ordering (Figure 4). Specifically, we find that σ = 0.1 (stretching along Γ-M and compressing along the perpendicular Γ-K direction) stabilizes the CDW-$4a_0$ order, with a peak in $D_2(\mathbf{q})$ forming between 0.25|$\mathbf{Q}_{Bragg}$| and 0.28|$\mathbf{Q}_{Bragg}$| momentum transfer wave vector (Fig. 4). The predicted CDW geometry is 3Q, but inclusion of anisotropy in the phonon energies of around 1.8%, the same order of magnitude as the strain, is enough to yield the experimentally observed 1Q state. Similarly, we find that σ = -0.3 (stretching along Γ-K and compressing along the perpendicular Γ-M direction) leads to a CDW with a peak in $D_2(\mathbf{q})$ forming near 0.4|$\mathbf{Q}_{Bragg}$| (Fig. 4). In this case, the energetic payoff of locking into the nearest commensurate structure [37], which is not included in the present model, would be expected to increase the CDW wavevector to the observed CDW-$2a_0$ period. While it is difficult to obtain the exact relationship between σ/$σ_i$ and the magnitude of real-space lattice distortion, the generic dependence of the orbital overlaps on interatomic distance found in for example Ref. [38], suggests that changes in the overlap integrals are expected to be approximately five times the relative strain as defined in the experimental analysis. Using this rough estimate, we calculate the magnitude and the direction of strain used in our model to achieve different CDWs, which leads to a reasonable agreement with the relative strain values observed in the experiment (SI Appendix, Section VI). Moreover, the electronic band dispersion used to calculate $D_2(\mathbf{q})$ in the presence of these strain levels presents a good match to the experimentally measured electronic dispersion obtained from the QPI data in Fig. 3. Remarkably, the calculations indicate that both 1Q and 3Q phases of CDW-$2a_0$ may be stabilized, which can in fact be observed in STM data acquired at higher bias (SI Appendix, Section VII).

Despite its simplicity, our model is able to reproduce the wave vectors and geometries of all observed CDWs, and points to the dominant physical mechanism behind the CDW formation. CDW order is sensitive to two effects of strain — softening of phonon energies and modification of electron hopping parameters — each playing a distinct role in the formation of the resulting CDW phase. The main effect of the changes in the phonon dispersion by strain is the favoring of one type of geometry ("stripe" 1Q) over another (triangular 3Q). The effect of the electronic modification, on the other hand, is to alter the CDW wavevector, and even relatively small strain can have a significant effect. Exploiting these trends, we should in principle be able to strain-engineer desired charge ordering structures in this and other materials by considering the shift in the peak in the electronic susceptibility.

Our simple platform for exerting strain on bulk single crystals presented here can be combined with a variety of characterization techniques. A single CDW domain can be found over microscopically large regions of the sample covering hundreds of nanometers (SI Appendix, Section VIII), so in addition to nanoscopic methods, micro-ARPES or micro-Raman spectroscopy could also be used to study these novel phases. Moreover, this strain technique can be applied to a range of other materials. For example, $1T$-TiSe$_2$ could be strained to induce superconductivity [39] or novel CDW wavevectors and geometries in analogy to what we observe in $2H$-NbSe$_2$. Similarly, Fe-based superconductors could be strained, potentially using substrates with a TEC along a preferred direction [3], to create a rich playground to study the interplay of nematic order and superconductivity [40] within a single material using SI-STM.

**Methods**

Single crystals of $2H$-NbSe$_2$ were grown using vapor transport growth technique with iodine (I$_2$) as the transport agent, and exhibit superconducting transition temperature $T_c$ ~ 7 K based on the onset of diamagnetic signal due to the Meissener effect in magnetization measurements (SI Appendix, Section IX). Superconducting transition temperature remained approximately the same with $T_c$ ~ 7 K after the samples were strained and re-measured. Typical size of the single crystals used was ~ 2 mm by 2 mm, with ~0.1 mm thickness before cleaving and ~0.01 mm to ~0.1 mm thickness post cleaving. Instead of attaching the $2H$-NbSe$_2$ crystals directly to a metallic holder with TEC comparable to that of NbSe$_2$, as typically used in most STM experiments, we use conducting epoxy (EPO-TEK H20E) to glue the bottom of NbSe$_2$ to silica (SiO$_2$), a material with a vastly different TEC (Fig. 1a). Then, the NbSe$_2$/silica structure is attached to the STM sample holder and cooled down to ~4.5 K (more information in SI Appendix, Section IX). Based on the difference between TECs of NbSe$_2$ and silica, NbSe$_2$ is expected to stretch isotopically in-plane by ~0.15%. As we demonstrate from STM topographs, the actual induced strain at the sample surface can be spatially inhomogeneous. To create a clean surface necessary for STM measurements, the samples were cleaved in UHV, and inserted into the STM head within minutes. We studied 4 different NbSe$_2$ crystals glued on silica (5 different surfaces as one sample was re-cleaved for the second approach). For each of these 5, we approached the tip on several different points on the sample, which are typically tens of micrometers away from one another, and searched for different types of CDWs. We observed: all three CDWs on 2 surfaces, just CDW-2a$_0$ and CDW-4a$_0$ on two other surfaces, and just CDW-3a$_0$ on one surface.

STM data was acquired using a Unisoku USM1300 STM at the base temperature of ~4.5 K. All spectroscopic measurements have been taken using a standard lock-in technique at 915 Hz frequency and varying bias excitation as detailed in the figure captions. The STM tips used were home-made, chemically etched W tips annealed to bright-orange color in UHV. Tip quality has been evaluated on the surface of single crystal Cu(111) prior to performing the measurements presented in this paper. The Cu(111) surface was cleaned by repeated cycles of heating and Argon sputtering in UHV before it was inserted into the STM head.

To construct a model which captures experimental observations, we employ a tight-binding fit to the ARPES data for the two bands crossing the Fermi level (described in detail in Refs. [26,35]). The model assumes the two bands to be bonding and antibonding combinations of the two Nb $d_{3z^2-r^2}$ orbitals. We include both biaxial and uniaxial in-plane strain by modifying the hopping integrals based on the assumption that overlap integrals are linearly dependent on displacement, with an equal prefactor for all overlaps. In modeling uniaxial strain, we assume that a tensile strain in one direction leads to a compressive strain in the perpendicular in-plane direction, conserving the volume of the unit cell. Then, we employ the Random Phase Approximation to calculate the phonon softening as seen in resonant inelastic X-ray scattering [20,36]. The CDW wavevector is identified as the first wavevector to soften to zero. By including nonlinear terms in a Landau free energy expression we are able to reveal whether the CDW geometry consists of stripes (1Q) or triangles (3Q) (see SI Appendix, Section VI for more details).

## Acknowledgements

We thank Peter Littlewood and Vidya Madhavan for helpful conversations. F.F. acknowledges support from a Lindemann Trust Fellowship of the English-Speaking Union, and the Astor Junior Research Fellowship from New College, Oxford. J.v.W. acknowledges support from a VIDI grant financed by the Netherlands Organisation for Scientific Research (NWO). K.S.B. appreciates support from the National Science Foundation under Grant Number NSF-DMR-1709987. Z.W. is supported by the DOE Grant No. DE-FG02-99ER45747. I.Z. gratefully acknowledges the support from the National Science Foundation under Grant Number NSF-DMR-1654041 for the partial support of S.G, H.Z, B.R. and Z.R.9

**References**

1. Hicks, C. W. *et al.* Strong increase of $T_c$ of $Sr_2RuO_4$ under both tensile and compressive strain. *Science* 344, 283–5 (2014).

2. Chu, J.-H. *et al.* In-plane resistivity anisotropy in an underdoped iron arsenide superconductor. *Science* 329, 824–6 (2010).

3. He, M. *et al.* Dichotomy between in-plane magnetic susceptibility and resistivity anisotropies in extremely strained $BaFe_2As_2$. *Nat. Commun.* 8, 504 (2017).

4. Manzeli, S., Ovchinnikov, D., Pasquier, D., Yazyev, O. V. & Kis, A. 2D transition metal dichalcogenides. *Nat. Rev. Mater.* 2, 17033 (2017).

5. Roldán, R., Castellanos-Gomez, A., Cappelluti, E. & Guinea, F. Strain engineering in semiconducting two-dimensional crystals. *J. Phys. Condens. Matter* 27, 313201 (2015).

6. Levy, N. *et al.* Strain-induced pseudo-magnetic fields greater than 300 tesla in graphene nanobubbles. *Science* 329, 544–7 (2010).

7. Zhu, S., Stroscio, J. A. & Li, T. Programmable Extreme Pseudomagnetic Fields in Graphene by a Uniaxial Stretch. *Phys. Rev. Lett.* 115, 245501 (2015).

8. Hicks, C. W., Barber, M. E., Edkins, S. D., Brodsky, D. O. & Mackenzie, A. P. Piezoelectric-based apparatus for strain tuning. *Rev. Sci. Instrum.* 85, 65003 (2014).

9. Wilson, J. a., Di Salvo, F. J. & Mahajan, S. Charge-density waves and superlattices in the metallic layered transition metal dichalcogenides. *Adv. Phys.* 24, 117–201 (1975).

10. Arguello, C. J. *et al.* Quasiparticle interference, quasiparticle interactions, and the origin of the charge density wave in 2H-$NbSe_2$. *Phys. Rev. Lett.* 114, 37001 (2015).

11. Berthier, C., Molinié, P. & Jérome, D. Evidence for a connection between charge density waves and the pressure enhancement of superconductivity in 2H-$NbSe_2$. *Solid State Commun.* 18, 1393–1395 (1976).

12. Borisenko, S. V. *et al.* Two energy gaps and fermi-surface 'arcs' in $NbSe_2$. *Phys. Rev. Lett.* 102, 166402 (2009).

13. Chatterjee, U. *et al.* Emergence of coherence in the charge-density wave state of 2H-$NbSe_2$. *Nat. Commun.* 6, 6313 (2015).

14. Feng, Y. *et al.* Itinerant density wave instabilities at classical and quantum critical points. *Nat. Phys.* 11, 865–871 (2015).

15. Feng, Y. *et al.* Order parameter fluctuations at a buried quantum critical point. *Proc. Natl. Acad. Sci.* 109, 7224–7229 (2012).

16. Harper, J. M. E., Geballe, T. H. & Di Salvo, F. J. Heat capacity of 2H-NbSe$_2$ at the charge density wave transition. *Phys. Lett. A* 54, 27–28 (1975).

17. Kiss, T. *et al.* Charge-order-maximized momentum-dependent superconductivity. *Nat. Phys.* 3, 720–725 (2007).

18. Hou, X.-Y. *et al.* Proximity-Induced Superconductivity in New Superstructures on 2H-NbSe$_2$ Surface. *Chinese Phys. Lett.* 34, 77403 (2017).

19. Rahn, D. J. *et al.* Gaps and kinks in the electronic structure of the superconductor 2H-NbSe$_2$ from angle-resolved photoemission at 1 K. *Phys. Rev. B* 85, 224532 (2012).

20. Weber, F. *et al.* Extended Phonon Collapse and the Origin of the Charge-Density Wave in 2H–NbSe$_2$. *Phys. Rev. Lett.* 107, 107403 (2011).

21. Soumyanarayanan, A. *et al.* Quantum phase transition from triangular to stripe charge order in NbSe$_2$. *Proc. Natl. Acad. Sci.* 110, 1623–1627 (2013).

22. Kundu, H. K. *et al.* Quantum Phase Transition in Few-Layer NbSe$_2$ Probed through Quantized Conductance Fluctuations. *Phys. Rev. Lett.* 119, 226802 (2017).

23. van Wezel, J., Nahai-Williamson, P. & Saxena, S. S. Exciton-phonon-driven charge density wave in TiSe$_2$. *Phys. Rev. B* 81, 165109 (2010).

24. Rice, T. M. & Scott, G. K. New Mechanism for a Charge-Density-Wave Instability. *Phys. Rev. Lett.* 35, 120–123 (1975).

25. Doran, N. J., Titterington, D. J., Ricco, B. & Wexler, G. A tight binding fit to the bandstructure of 2H-NbSe$_2$ and NbS$_2$. *J. Phys. C Solid State Phys.* 11, 685–698 (1978).

26. Flicker, F. & van Wezel, J. Charge order from orbital-dependent coupling evidenced by NbSe$_2$. *Nat. Commun.* 6, 7034 (2015).

27. Rosenthal, E. P. *et al.* Visualization of electron nematicity and unidirectional antiferroic fluctuations at high temperatures in NaFeAs. *Nat. Phys.* 10, 225–232 (2014).

28. Giambattista, B., Johnson, A., Coleman, R. V., Drake, B. & Hansma, P. K. Charge-density waves observed at 4.2 K by scanning-tunneling microscopy. *Phys. Rev. B* 37, 2741–2744 (1988).

29. Flicker, F. & van Wezel, J. Charge ordering geometries in uniaxially strained NbSe$_2$. *Phys. Rev. B* 92, 201103 (2015).

30. Lawler, M. J. *et al.* Intra-unit-cell electronic nematicity of the high-$T_c$ copper-oxide pseudogap states. *Nature* 466, 347–351 (2010).

31. Liu, Y. *et al.* Tuning Dirac states by strain in the topological insulator $Bi_2Se_3$. *Nat. Phys.* 10, 294–299 (2014).

32. Zeljkovic, I. *et al.* Strain engineering Dirac surface states in heteroepitaxial topological crystalline insulator thin films. *Nat. Nanotechnol.* 10, 849–853 (2015).

33. Hÿtch, M. J., Snoeck, E. & Kilaas, R. Quantitative measurement of displacement and strain fields from HREM micrographs. *Ultramicroscopy* 74, 131–146 (1998).

34. Zhao, J. *et al.* Orbital selectivity causing anisotropy and particle-hole asymmetry in the charge density wave gap of 2H–$TaS_2$. *Phys. Rev. B* 96, 125103 (2017).

35. Flicker, F. & van Wezel, J. Charge order in NbSe2. *Phys. Rev. B* 94, 235135 (2016).

36. Weber, F. *et al.* Optical phonons and the soft mode in 2H-$NbSe_2$. *Phys. Rev. B* 87, 245111 (2013).

37. McMillan, W. L. Landau theory of charge-density waves in transition-metal dichalcogenides. *Phys. Rev. B* 12, 1187–1196 (1975).

38. Wills, J. M. & Harrison, W. A. Interionic interactions in transition metals. *Phys. Rev. B* 28, 4363–4373 (1983).

39. Joe, Y. I. *et al.* Emergence of charge density wave domain walls above the superconducting dome in 1T-$TiSe_2$. *Nat. Phys.* 10, 421–425 (2014).

40. Chubukov, A. V., Khodas, M. & Fernandes, R. M. Magnetism, Superconductivity, and Spontaneous Orbital Order in Iron-Based Superconductors: Which Comes First and Why? *Phys. Rev. X* 6, 41045 (2016).

41. Moncton, D. E., Axe, J. D. & DiSalvo, F. J. Neutron scattering study of the charge-density wave transitions in 2H–$TaSe_2$ and 2H–$NbSe_2$. *Phys. Rev. B* 16, 801–819 (1977).

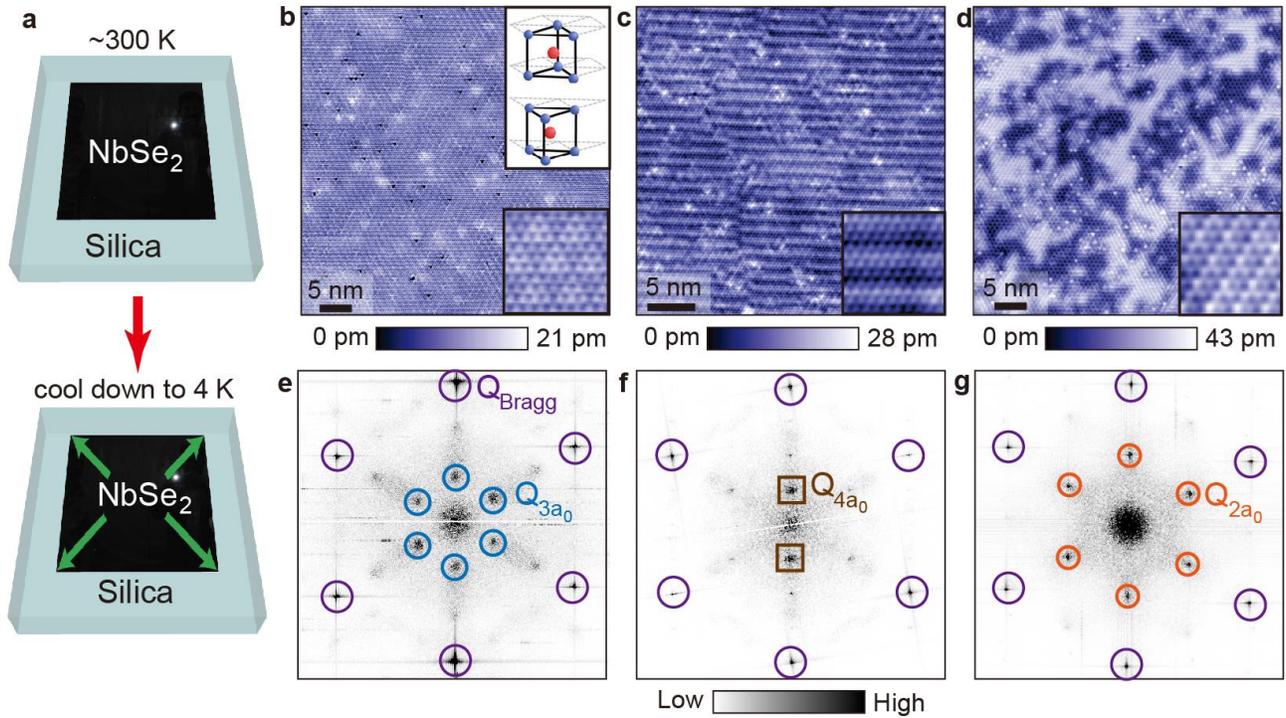

**Figure 1. Strain generation method and induced charge density wave phases in *2H*-NbSe$_2$.** (a) An illustration of how strain is applied to the sample. A single crystal is attached to the top of a silica plate by silver epoxy at room temperature. When the sample and the substrate are both cooled down to ~4 K, the difference of the TECs between the two materials will cause NbSe$_2$ to stretch. (b-d) STM topographs and (e-g) their Fourier transforms of CDW-3a$_0$, CDW-4a$_0$ and CDW-2a$_0$ regions of the sample, respectively. Atomic Bragg peaks in (e-g) are circled in purple, while the CDW peaks corresponding to CDW-3a$_0$, CDW-4a$_0$ and CDW-2a$_0$ are enclosed in blue, brown and orange, respectively. The top right inset in (b) shows the crystal structure of *2H*-NbSe$_2$ (Se atoms shown in blue and Nb atoms in red). The bottom insets in (b-d) show close-ups on each CDW phase. STM setup conditions are: (b) $I_{set}$ = 42 pA and $V_{sample}$ = -200 mV; (c) $I_{set}$ = 200 pA and $V_{sample}$ = 60 mV and (d) $I_{set}$ = 500 pA and $V_{sample}$ = -200 mV.

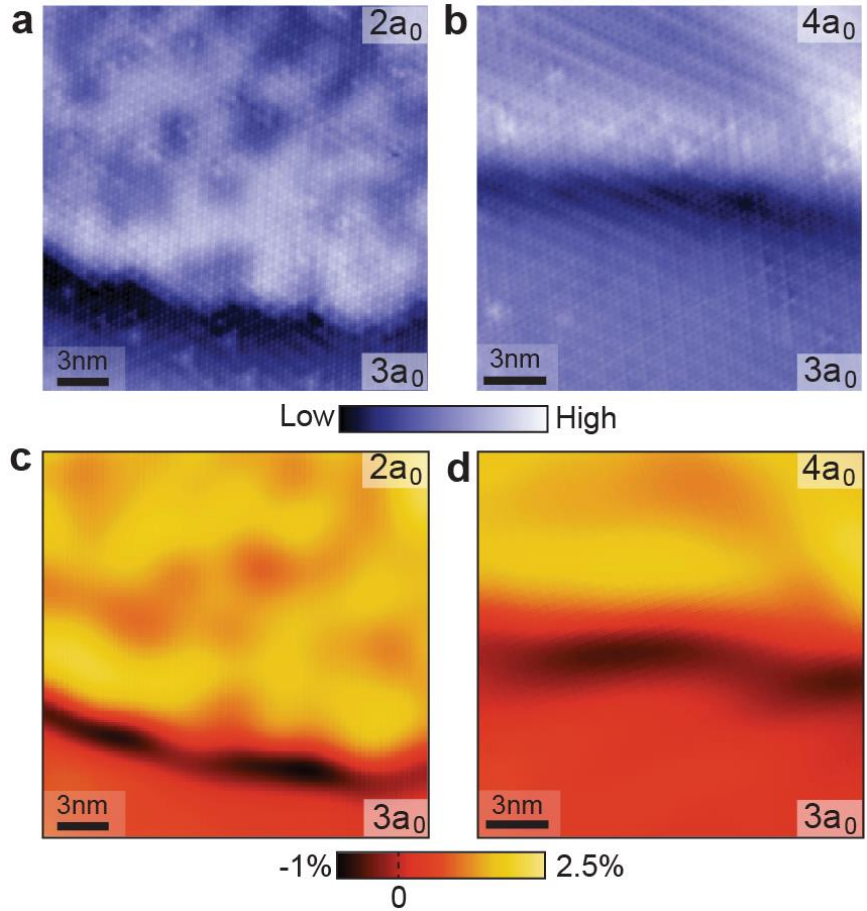

**Figure 2. Local strain mapping.** (a,b) STM topographs and (c,d) biaxial (isotropic) strain maps of the atomically-smooth boundaries between regions hosting different CDW phases. The biaxial strain maps have been calculated from the derivatives of the strain fields as $(s_{xx} + s_{yy})/2$, using the procedure described in SI Appendix, Section II. The algorithm assumes that strain is zero in the CDW-$3a_0$ area, and calculates the relative strain with respect to it. Larger positive values represent tensile strain (stretching of the lattice). As can be seen, both CDW-$2a_0$ and CDW-$4a_0$ regions are characterized by tensile strain relative to the CDW-$3a_0$ area. STM setup conditions were: (a) $I_{set}$ = 350 pA and $V_{sample}$ = -70 mV; (b) $I_{set}$ = 200 pA and $V_{sample}$ = -100 mV.

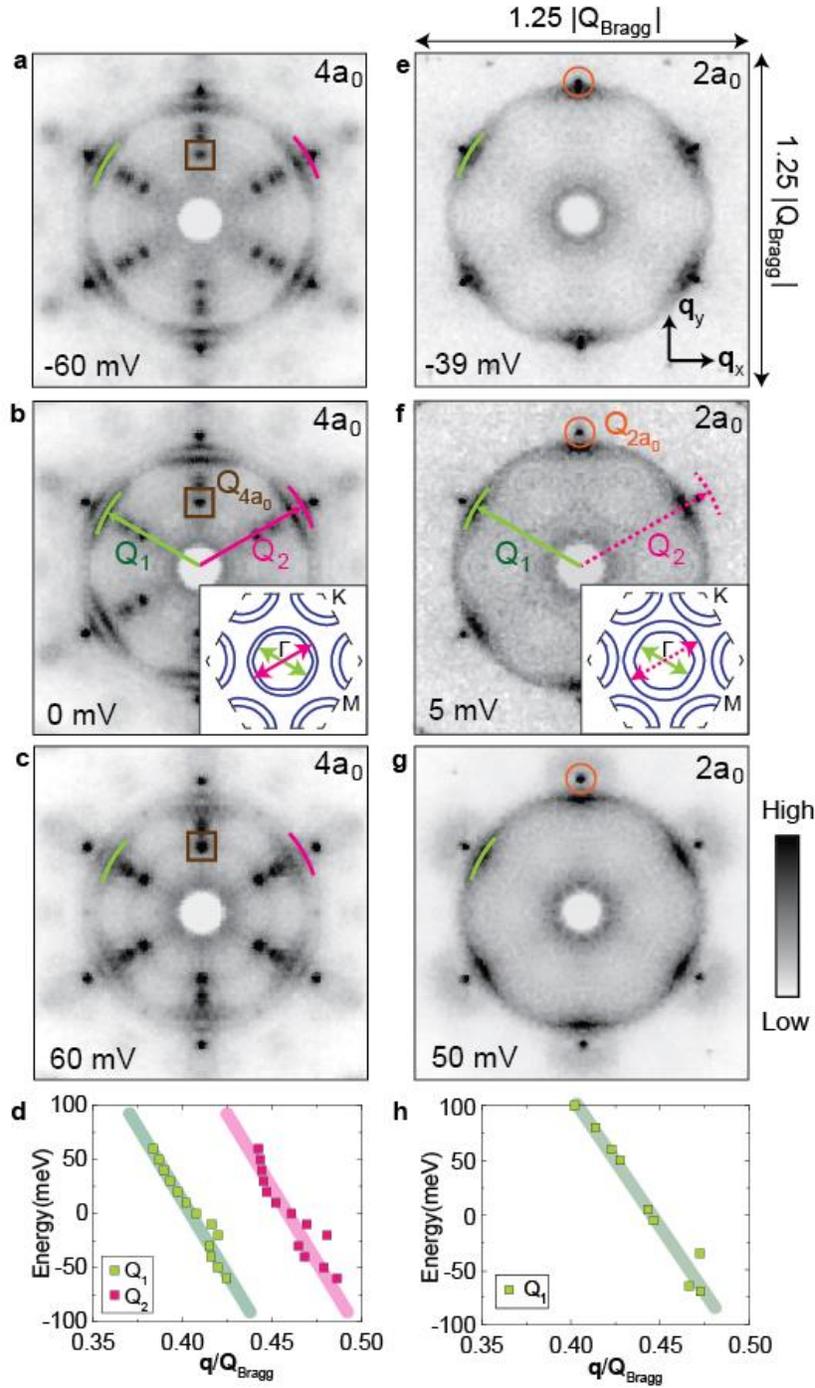

**Figure 3. Electronic band structure mapping using QPI imaging.** Fourier transforms (FTs) of *dI/dV* maps acquired at (a) -60 mV, (b) 0 mV and (c) 60 mV over a CDW-$4a_0$ region of the sample. The inset in (b) shows the schematic of the Fermi surface within the first Brillouin zone. (d) The dispersion of the QPI peaks as a function of energy along the Γ-M direction in the CDW-$4a_0$ region. FTs of *dI/dV* maps acquired at (e) -39 mV, (f) 5 mV and (g) 50 mV over the CDW-$2a_0$ region of the sample. The inset in (f) shows the schematic of the Fermi surface under small tensile strain, which is expected to move the Fermi surface pockets around Γ further apart. Only $Q_1$ vector in (e-g) can be seen, while $Q_2$ is notably absent. (h) The dispersion of the QPI peak as a function of energy along the Γ-M direction in the CDW-$2a_0$ region. QPI peak positions in (d,h) are determined using Gaussian peak fitting to a one-dimensional curve extracted along a line connecting the center of the FT and the atomic Bragg peak. QPI peaks and CDW peaks are denoted by the guides for the eye in panels (a-c,e-g): $Q_1$ (green line), $Q_2$ (pink line), $Q_{2a0}$ (orange circle) and $Q_{4a0}$ (brown square). The center of all FTs has been artificially suppressed to emphasize other features. All FTs have been six-fold symmetrized to enhance signal-to-noise, and cropped to the same $1.25|Q_{Bragg}|$ square size window. The region of the sample where the data in (a-c) was taken contains domains of CDW-$4a_0$ along only two lattice directions (Fig. S2(a)). As CDW-$4a_0$ is intrinsically a unidirectional order, the six-fold symmetry of the $Q_{4a0}$ peak in (a-c) is an artifact of the symmetrization process. STM setup conditions: (a-c) $I_{set}$ = 320 pA, $V_{sample}$ = -60 mV and $V_{exc}$ = 10 mV (zero-to-peak); (e) $I_{set}$ = 200 pA, $V_{sample}$ = -39 mV and $V_{exc}$ = 1 mV; (f) $I_{set}$ = 20 pA, $V_{sample}$ = 5 mV and $V_{exc}$ = 1.5 mV; (g) $I_{set}$ = 300 pA, $V_{sample}$ = 50 mV and $V_{exc}$ = 10 mV.

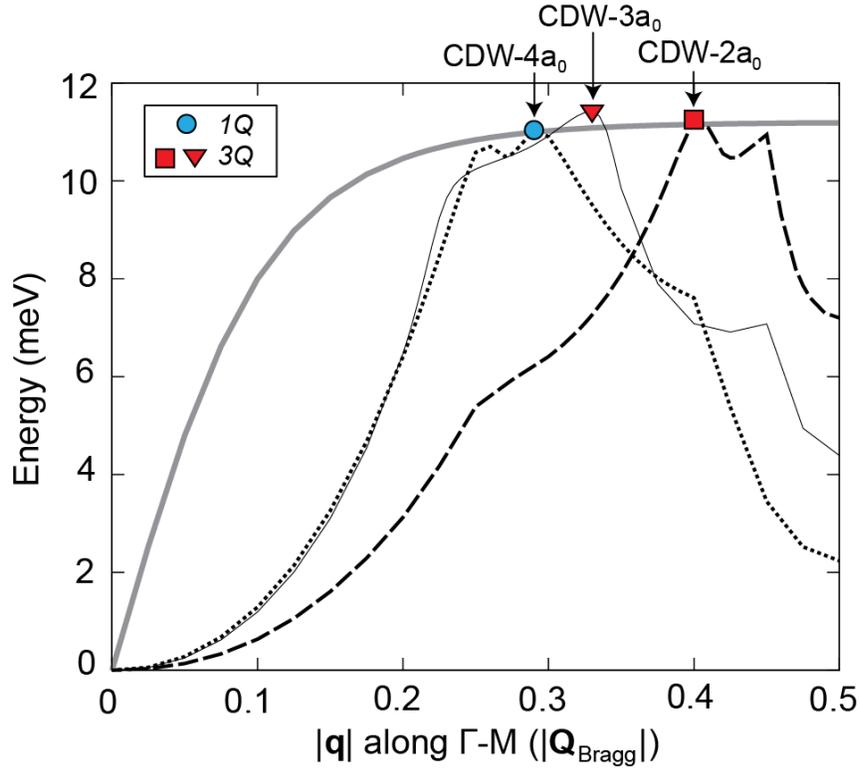

**Figure 4. Theoretical modeling.** The dispersion $\Omega_0$ of the longitudinal acoustic phonons extracted from experimental RIXS data [20] (thick solid grey line). All other curves and symbols represent results from our theoretical simulation, which computes the electronic susceptibility $D_2(\mathbf{q})$ as a function of strain. Following Ref. [35], $D_2(\mathbf{q})$ is defined in meV, so that the charge order is expected to develop whenever $D_2$ exceeds $\Omega_0$. The thin solid black line represents $D_2(\mathbf{q})$ for the unstrained case with 3Q order and wavevector ~ $0.33|\mathbf{Q}_{Bragg}|$, in agreement with the observed value [41]. The dotted line represents $D_2(\mathbf{q})$ for a uniaxial stretch along Γ-M (modeled by $\sigma = 0.1$) resulting in a 1Q CDW-$4a_0$ with wavevector ~$0.25|\mathbf{Q}_{Bragg}|$. The thick dashed line represents $D_2(\mathbf{q})$ for uniaxial strain in the perpendicular direction (modeled by $\sigma = -0.3$) resulting in a peak at ~$0.4|\mathbf{Q}_{Bragg}|$. In practice, this will most likely result in locking into a commensurate CDW-$2a_0$ with $0.5|\mathbf{Q}_{Bragg}|$ wave vector when lattice-interaction effects are included [37]. Red (blue) symbols indicate the first point to order into a 3Q (1Q) CDW geometry. Following the generic considerations of Ref. [38], the predicted relative changes in orbital overlap may be expected to be roughly five times the relative strain defined in the experimental analysis, as explained in the Supplementary Material VI.